\documentclass[11pt,letterpaper]{article}
\usepackage{amsfonts,color,ifpdf,latexsym,graphicx,subfigure,url,verbatim}
\title{(Non)existence of Pleated Folds: \\
       How Paper Folds Between Creases}
\author{%
  Erik D. Demaine%
    \thanks{MIT Computer Science and Artificial Intelligence Laboratory,
      32 Vassar St., Cambridge, MA 02139, USA,
      \protect\url{{edemaine,mdemaine,price}@mit.edu}.}
    \thanks{Partially supported by NSF CAREER award CCF-0347776.}
\and
  Martin L. Demaine%
    \footnotemark[1]
\and
  Vi Hart%
    \thanks{\protect\url{http://vihart.com}}
\andlinebreak
  Gregory N. Price%
    \footnotemark[1]
\and
  Tomohiro Tachi%
    \thanks{Department of Architecture, The University of Tokyo,
      7-3-1 Hongo, Bunkyo, Tokyo 113-8658, Japan,
      \protect\url{ttachi@siggraph.org}}
}
\date{}

\newif\ifabstract
\abstracttrue
\newif\iffull
\ifabstract \fullfalse \else \fulltrue \fi

\usepackage
  [breaklinks,bookmarks,bookmarksnumbered,bookmarksopen,bookmarksopenlevel=2]
  {hyperref}
{\makeatletter \hypersetup{pdftitle={\@title}}}
\hypersetup{pdfauthor={Erik D. Demaine, Martin L. Demaine, Vi Hart, Gregory N. Price, and Tomohiro Tachi}}

\advance \topmargin by -\headheight
\advance \topmargin by -\headsep
\evensidemargin \oddsidemargin
{\makeatletter
 \gdef\xxxmark{%
   \expandafter\ifx\csname @mpargs\endcsname\relax 
     \expandafter\ifx\csname @captype\endcsname\relax 
       \marginpar{xxx}
     \else
       xxx 
     \fi
   \else
     xxx 
   \fi}
 \gdef\xxx{\@ifnextchar[\xxx@lab\xxx@nolab}
 \long\gdef\xxx@lab[#1]#2{\textbf{[\xxxmark #2 ---{\sc #1}]}}
 \long\gdef\xxx@nolab#1{\textbf{[\xxxmark #1]}}
}

{\makeatletter \gdef\fps@figure{!htbp}}
{\makeatletter \gdef\fps@table{!htbp}}

\def\andlinebreak{\end{tabular}\linebreak\begin{tabular}[t]{c}}

\let\realbfseries=\bfseries
\def\bfseries{\realbfseries\boldmath}

\newtheorem{theorem}{Theorem}
\newtheorem{lemma}[theorem]{Lemma}
\newtheorem{proposition}[theorem]{Proposition}
\newtheorem{corollary}[theorem]{Corollary}
\newtheorem{definition}{Definition}

\newenvironment{proof}{\noindent\textbf{Proof: }\ignorespaces}
  {\hspace*{\fill}$\Box$\medskip}

\let\epsilon=\varepsilon
\newcommand\eps{\epsilon}
\newcommand\term[1]{\emph{#1}}
\newcommand\R{\mathbb R}
\newcommand\interior{\mathop{\rm interior}}

\begin{document}
\maketitle

\begin{abstract}
  We prove that the pleated hyperbolic paraboloid, a familiar origami
  model known since 1927, in fact cannot be folded with the standard crease
  pattern in the standard mathematical model of zero-thickness paper.
  In contrast, we show that the model can be folded with additional creases,
  suggesting that real paper ``folds'' into this model via small such creases.
  We conjecture that the circular version of this model, consisting simply
  of concentric circular creases, also folds without extra creases.

  At the heart of our results is a new structural theorem characterizing
  uncreased intrinsically flat surfaces---the portions of paper between
  the creases.  Differential geometry has much to say about the local
  behavior of such surfaces when they are sufficiently smooth, e.g.,
  that they are torsal ruled.  But this classic result is simply false
  in the context of the whole surface.  Our structural characterization
  tells the whole story, and even applies to surfaces with discontinuities
  in the second derivative.
  We use our theorem to prove fundamental properties about how paper folds,
  for example, that straight creases on the piece of paper
  must remain piecewise-straight (polygonal) by folding.
\end{abstract}


\section{Introduction}

\begin{figure}
  \centering
  \subfigure[Standard mountain-valley pattern.
             \label{hypar cp} \label{hypar mv}]
            {\includegraphics[scale=0.6]{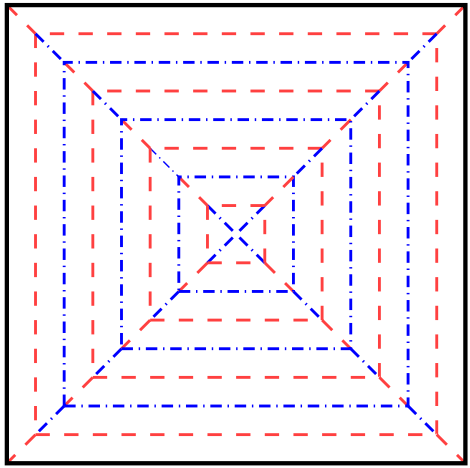}}\hfil
  \subfigure[Photograph of physical model. {[Jenna Fizel]} \label{hypar photo}]
            {\includegraphics[scale=0.2]{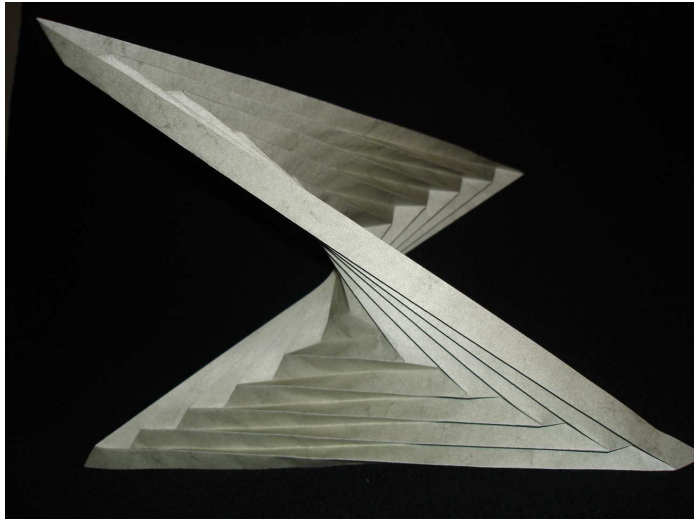}}
  \caption{Pleated hyperbolic paraboloid.}
  \label{hypar}
\end{figure}

\begin{figure}
  \centering
  \subfigure[Mountain-valley pattern.
             \label{circular cp} \label{circular mv}]
            {\includegraphics[scale=0.6]{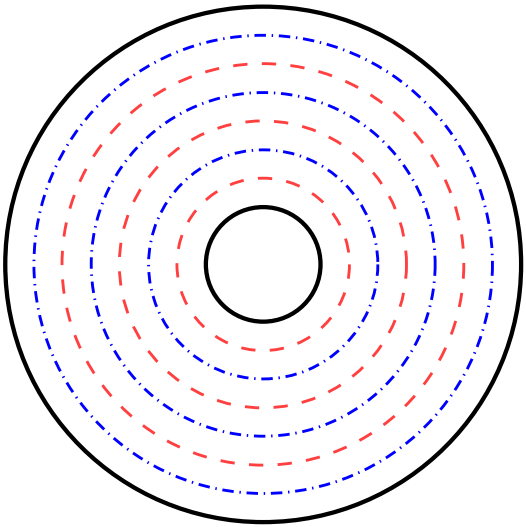}}\hfil
  \subfigure[Photograph of physical model. {[Jenna Fizel]} \label{circular photo}]
            {\includegraphics[scale=0.2,trim=100 0 0 0,clip]{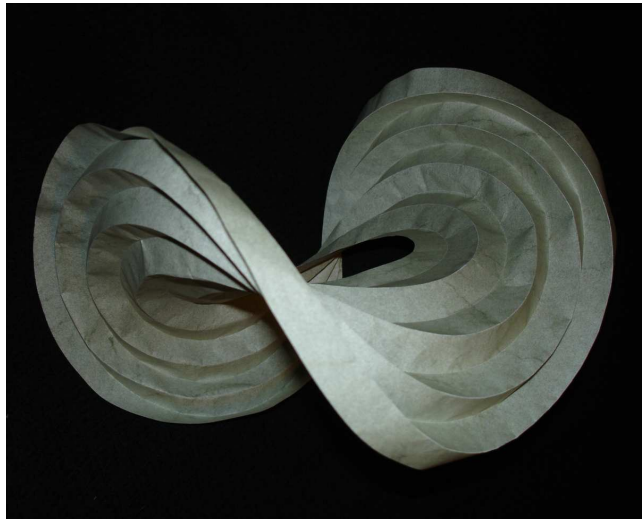}}
  \caption{Circular pleat.}
  \label{circular}
\end{figure}

A fascinating family of \emph{pleated} origami models use extremely simple
crease patterns---repeated concentric shapes, alternating mountain and
valley---yet automatically fold into interesting 3D shapes.
The most well-known is the \emph{pleated hyperbolic paraboloid},
shown in Figure~\ref{hypar}, where the crease pattern is concentric squares
and their diagonals.
As the name suggests, it has long been conjectured, but never formally
established, that this model approximates a hyperbolic paraboloid.
More impressive (but somewhat harder to fold) is the \emph{circular pleat},
shown in Figure~\ref{circular}, where the crease pattern is simply
concentric circles, with a circular hole cut out of the center.
Both of these models date back to the Bauhaus, from a preliminary course in
paper study taught by Josef Albers in 1927--1928 \cite[p.~434]{Wingler-1969},
and taught again later at Black Mountain College in 1937--1938
\cite[pp.~33, 73]{Adler-2004}; see \cite{curved}.
These models owe their popularity today to origamist Thoki Yenn,
who started distributing the model sometime before 1989.
Examples of their use and extension for algorithmic sculpture
include \cite{BRIDGES99,AAG08}.

The magic of these models is that most of the actual folding happens by
the physics of paper itself; the origamist simply puts all the creases in
and lets go.  Paper is normally elastic: try wrapping a paper sheet
around a cylinder, and then letting go---it returns to its original state.
But \emph{creases} plastically deform the paper beyond its yield point,
effectively resetting the elastic memory of paper to a nonzero angle.
Try creasing a paper sheet and then letting go---it stays folded at the crease.
The harder you press the crease, the larger the desired fold angle.
What happens in the pleated origami models is that the paper tries to
stay flat in the uncreased portions, while trying to stay folded at the
creases, and physics computes a configuration that balances these forces
in equilibrium (with locally minimum free energy).

But some mathematical origamists have wondered over the years \cite{nytimes}:
do these models actually \emph{exist}?  Is it really possible to fold
a piece of paper along exactly the creases in the crease pattern of
Figures~\ref{hypar} and~\ref{circular}?
The first two authors have always suspected that both models existed,
or at least that one existed if and only if the other did.
But we were wrong.

\paragraph{Our results.}
We prove that the hyperbolic-paraboloid crease pattern of Figure~\ref{hypar cp}
does not fold using exactly the given creases,
even with a hole cut out of the center.

In proving the impossibility of folding the pleated hyperbolic
paraboloid, we develop a structural characterization of how uncreased
paper can fold (hence the title of this paper).  Surprisingly, such a
characterization has not been obtained before.  An intuitive
understanding (often misquoted) is that paper folds like a ruled
surface, but that claim is only true locally (infinitesimally) about
every point.  When the paper is not smooth or has zero principal
curvature at some points, the truth gets even subtler.  We correct
both of these misunderstandings by handling nonsmooth (but uncreased)
surfaces, and by stating a local structure theorem flexible enough to
handle zero curvatures and all other edge cases of uncreased surfaces.

In contrast, we conjecture that the circular-pleat crease pattern of
Figure~\ref{circular cp} folds using exactly the given creases, when
there is a hole cut out of the center.  A proof of this would be the
first proof of existence of a curved-crease origami model (with more
than one crease) of which we are aware.  Existing work characterizes
the local folding behavior in a narrow strip around a curved crease,
and the challenge is to extend this local study to a globally consistent
folding of the entire crease pattern.

Another natural remaining question is what actually happens to a real
pleated hyperbolic paraboloid like Figure~\ref{hypar photo}.
One conjecture is that the paper uses extra creases (discontinuities
in the first derivative), possibly many very small ones.
We prove that, indeed, simply triangulating the crease pattern,
and replacing the four central triangles with just two triangles,
results in a foldable crease pattern.
Our proof of this result is quite different in character,
in that it is purely computational instead of analytical.
We use interval arithmetic to establish with certainty that the
exact object exists for many parameter values, and its coordinates
could even be expressed by radical expressions in principle,
but we are able only to compute arbitrarily close approximations.


\section{Structure of Uncreased Flat Surfaces}

Our impossibility result rests on an understanding of how it is possible
to fold the faces of the crease pattern, which by definition are regions
folded without creases.  The geometric crux of the proof therefore relies
on a study of uncreased intrinsically flat (paper) surfaces.
This section gives a detailed analysis of such surfaces.
Our analysis allows discontinuities all the way down to the second derivative
(but not the first derivative---those are creases),
provided those discontinuities are somewhat tame.

We begin with some definitions, in particular to nail down the notion
of creases.

\begin{definition}
  For us a \term{surface} is a compact 2-manifold embedded in $\R^3$.
  The surface is $C^k$ if the manifold and its embedding are $C^k$.
  The surface is \term{piecewise-$C^k$} if it can be decomposed as a
  complex of $C^k$ regions joined by vertices and $C^k$ edges.
\end{definition}

\begin{definition}
  A \term{good surface} is a piecewise-$C^2$ surface.  A good surface
  $S$ therefore decomposes into a union of $C^2$ surfaces $S_i$,
  called \term{pieces}, which share $C^2$ edges $\gamma_j$, called
  \term{semicreases}, whose endpoints are \term{semivertices}.
  Isolated points of $C^2$ discontinuities are also \term{semivertices}.
  If $S$ is itself $C^1$ everywhere on a semicrease,
  we call it a \term{proper semicrease}; otherwise it is a \term{crease}.
  Similarly a semivertex $v$ is a \term{vertex} if $S$ is not $C^1$ at~$v$.
  Accordingly an \term{uncreased surface} is a $C^1$ good surface
  (with no creases or vertices),
  and a \term{creased surface} is a good surface not everywhere~$C^1$
  (with at least one crease or vertex).
\end{definition}

\begin{definition}
  A surface is \term{(intrinsically) flat} if every point $p$ has a
  neighborhood isometric to a region in the plane.%
  \footnote{Henceforth we use the term ``flat'' for this intrinsic notion
    of the surface metric, and the term ``planar'' for the extrinsic notion
    of (at least locally) lying in a 3D plane.}
\end{definition}

In order to understand the uncreased flat surfaces that are our chief
concern, we study the $C^2$ flat surfaces that make them up.
On a $C^2$ surface, the well-known \term{principal curvatures}
$\kappa_1 \geq \kappa_2$ are defined for each interior point as the
maximum and minimum (signed) curvatures for geodesics through the
point.  A consequence of Gauss's celebrated Theorema Egregium
\cite{Gauss-1902} is that, on a $C^2$ flat surface, the \term{Gaussian
  curvature} $\kappa_1\kappa_2$ must be everywhere zero.  Thus every
interior point of a $C^2$ flat surface is either \term{parabolic} with
$k_2 \neq k_1 = 0$ or \term{planar} with $k_2 = k_1 = 0$.

Each interior point $p$ on a $C^2$ flat surface therefore either
\begin{enumerate}
\item[(a)] is planar, with a planar neighborhood;
\item[(b)] is planar and the limit of parabolic points; or
\item[(c)] is parabolic, and has a parabolic neighborhood by continuity,
\end{enumerate}
and an interior point on an uncreased flat surface may additionally
\begin{enumerate}
\item[(d)] lie on the interior of a semicrease; or
\item[(e)] (a priori) be a semivertex.
\end{enumerate}

For points of type (a), it follows by integration that the
neighborhood has a constant tangent plane and indeed lies in this
plane.  Types (b) and (c) are a bit more work to classify, but the
necessary facts are set forth by
Spivak \cite[vol.~3, chap.~5, pp.~349--362]{Spivak-1979}
and recounted below.
(In Spivak's treatment the regularity condition is left unspecified,
but the proofs go through assuming only~$C^2$.)
We address type (d) farther below.  From our results it will become clear that
the hypothetical type (e) does not occur in uncreased flat surfaces.

\begin{proposition}
  {\rm \cite[Proposition III.5.4 et seq.]{Spivak-1979}} \label{prp:1}
  For every point $p$ of type (c) on a surface~$M$, a
  neighborhood $U \subset M$ of $p$ may be parametrized as
  $$ f(s,t) = c(s) + t \cdot \delta(s) $$
  where $c$ and $\delta$ are $C^1$ functions; $c(0) = p$;
  $|\delta(s)| = 1$; $c'(s)$, $\delta(s)$, and $\delta'(s)$ are coplanar
  for all $s$; and every point of $U$ is parabolic.
\end{proposition}

Write $\interior(M)$ for the interior of a surface~$M$.

\begin{proposition}
  {\rm \cite[Corollaries III.5.6--7]{Spivak-1979}} \label{prp:2}
  For every point $p$ of type (b) or (c) on a surface~$M$,
  there is a unique line $L_p$ passing through $p$ such that
  the intersection $L_p \cap M$ is open in $L_p$ at~$p$.
  The component $C_p$ containing $p$ of the
  intersection $L_p \cap \interior(M)$ is an open segment,
  and every point in $C_p$ is also of type (b) or (c) respectively.
\end{proposition}

Following the literature on flat surfaces, we speak of a segment like
the $C_p$ of Proposition~\ref{prp:2} as a \term{rule segment}.
The \term{ruling} of a surface is the family of rule segments of all
surface points, whose union equal the surface.
A ruling is \term{torsal} if all points along each rule segment
have a common tangent plane.

To characterize points of type (d), lying on semicreases, we require
the following two propositions.

\begin{proposition}\label{prp:6}
  Consider a point $q$ of type (d) on a surface~$M$.
  Then $q$ is not the endpoint of the rule segment $C_p$
  for any point $p \in M$ of type (b) or (c).
\end{proposition}
\begin{proof}
  It suffices to show the conclusion for $p$ of type (c), because a
  rule segment of type (b) is a limit of rule segments of type (c).
  Let $\gamma$ be the interior of the semicrease on which $q$ lies.

  Because $M$ is $C^1$, it has a tangent plane $M_q$ at each $q$, which
  is common to the two $C^2$ pieces bounded by $\gamma$.  Parametrize
  $\gamma$ by arclength with $\gamma(0) = q$, and write $n(s)$ for the
  unit normal to the tangent plane $M_{\gamma(s)}$.  Parametrize the
  two pieces as torsal ruled surfaces by the common curve $c_1(s) =
  c_2(s) = \gamma(s)$ and lines $\delta_1(s)$ and $\delta_2(s)$.
  Then, because each piece is torsal, $\dot n(s) \perp \delta_1(s)$ and
  $\dot n(s) \perp \delta_2(s)$.  But both $\delta_1(s)$ and $\delta_2(s)$
  lie in the tangent plane at $s$, perpendicular to $n(s)$, and so too
  does $\dot n(s)$ because $n(s)$ is always a unit vector.  Therefore,
  for each $s$, either $\dot n(s) = 0$ or $\delta_1(s)$ and $\delta_2(s)$
  are collinear.

  Let $A$ be the subset of $\gamma$ on which $\dot n(s) = 0$, and $B$
  the subset on which $\delta_1(s)$ and $\delta_2(s)$ are collinear.
  Then we have shown that $A \cup B = \gamma$.  By continuity, both
  $A$ and $B$ are closed.  Therefore any point of $\gamma$ which does
  not belong to $A$ is in the interior of $B$, and any point not in
  the interior of $A$ is in the closure of the interior of $B$.

  If an open interval $I$ along $\gamma$ is contained in $A$ so that
  $\dot n(s) = 0$, then a neighborhood in $M$ of $I$ is planar by
  integration because each rule segment has a single common tangent
  plane in a torsal ruled surface.  On the other hand if $I$ is
  contained in $B$ so that $\delta_1(s)$ and $\delta_2(s)$ are
  collinear, then a neighborhood is a single $C^2$ ruled surface.  In
  either case, the rule segments from one surface that meet $I$
  continue into the other surface.  That is, each rule segment meeting
  a point in the interior of $A$ or $B$ continues into the other
  surface.

  Now we conclude.  By continuity, each rule segment meeting a point
  in the closure of the interior of $A$ or $B$ continues into the
  other surface; but these two closures cover $\gamma$.  So no rule
  segment ends on $\gamma$, including at $q$.
\end{proof}

\begin{proposition}\label{prp:3}
  For every point $p$ of type (d) on a surface~$M$,
  there is a unique line $L_p$ passing through $p$
  such that the intersection $L_p \cap M$ is open in $L_p$ at~$p$.
  The component $C_p$ containing $p$ of the
  intersection $L_p \cap \interior(M)$ is an open segment,
  the limit of rule segments through points neighboring $p$,
  and every point of $C_p$ is also of type~(d).
\end{proposition}
\begin{proof}
  Let $B_r(p)$ be a radius-$r$ disk in $M$ centered at $p$,
  small enough that no point of the disk is a semivertex.
  By Proposition~\ref{prp:6}, the rule segment
  through any point $q$ of type (b) or (c) in the
  half-size disk $B_{r/2}(p)$ cannot end in $B_r(p)$,
  so it must be of length at least $r/2$ in each direction.
  Further, $p$ must be a limit of such points,
  or else a neighborhood of $p$ would be planar.

  By a simple compactness argument, provided in \cite{Spivak-1979} for the
  type-(b) case of Proposition~\ref{prp:2}, $C_p$ is the limit of (a
  subsequence of) rule segments $C_q$ through points $q$ of type (b)
  and (c) approaching $p$ and is an open segment.  Because each $C_q$
  has a single tangent plane, the discontinuity in second derivatives
  found at $p$ is shared along $C_p$.
\end{proof}

Two corollaries follow immediately from Proposition~\ref{prp:3}.

\begin{corollary}\label{cory:2}
  Every (proper) semicrease in an uncreased flat surface is a line segment,
  and its endpoints are boundary points of the surface.
\end{corollary}

\begin{corollary}\label{cory:3}
  An uncreased flat surface has no interior semivertices;
  every interior point is in the interior of a $C^2$ piece or a semicrease.
\end{corollary}

Another corollary summarizes much of Propositions~\ref{prp:2}
and~\ref{prp:3} combined.

\begin{corollary}\label{cory:1}
  Every interior point $p$ of an uncreased flat surface $M$ not
  belonging to a planar neighborhood belongs to a unique rule segment
  $C_p$.  The rule segment's endpoints are on the boundary of $M$, and
  every interior point of $C_p$ is of the same type (b), (c), or (d).
\end{corollary}

Finally, we unify the treatment of all types of points in the
following structure theorem for uncreased flat surfaces.  The theorem
is similar to Proposition~\ref{prp:1}, which concerns only points of
type~(c).

\begin{theorem}
  Every interior point of an uncreased flat surface has a neighborhood that is
  a ruled surface.  In each rule segment, every interior point is of the same
  type (a), (b), (c), or (d).  The ruled surface may be parametrized
  as
  $$ f(s, t) = c(s) + t \cdot \delta(s), $$
  where $c$ is $C^1$, $\delta$ is $C^0$, and $\delta$ is $C^1$ whenever
  $c(s)$ is of type (a), (b), or (c).
\end{theorem}
\begin{proof}
  Let $p$ be a point on an uncreased flat surface $M$.  If $p$ is of
  type (a), then we may parametrize its planar neighborhood as a ruled
  surface almost arbitrarily.  Otherwise, $p$ is of type (b), (c), or
  (d) and has a unique rule segment $C_p$.

  Embed a neighborhood $U \subset M$ of $C_p$ isometrically in the
  plane, by a map $\phi : U \to \R^2$.  Let $\gamma$ be a line segment
  in the plane perpendicularly bisecting $\phi(C_p)$, parametrized by
  arclength with $\gamma(0) = \phi(p).$  Every point $\phi^{-1}(\gamma(s))$
  of type (b), (c), or (d) has a unique rule segment $C_{\phi^{-1}(\gamma(s))}$;
  for such $s$, let $\eps(s)$ be the unit vector pointing along
  $\phi(C_{\phi^{-1}(\gamma(s))})$, picking a consistent orientation.

  Now the remaining $s$ are those for which $\phi^{-1}(\gamma(s))$ is of type
  (a).  These $s$ form an open subset, so that for each such $s$ there
  is a previous and next $s$ not of type (a).  For each such~$s$,
  we can determine an $\eps(s)$ by interpolating angles linearly between
  the $\eps(s)$ defined for the previous and next $s$ not of type (a).
  The resulting function $\eps(s)$ is continuous and identifies a
  segment through every point in $\gamma$, giving a parametrization of
  a neighborhood of $\gamma$ as a ruled surface by $g(s, t) =
  \gamma(s) + t \cdot \eps(s)$.

  Finally, write $f(s, t) = \phi^{-1}(g(s, t))$ to complete the construction.
\end{proof}


\section{How Polygonal Faces Fold}

If all edges of the crease pattern are straight,
every face of the crease pattern is a polygon.
We first show that, if the edges of such a polygon remain
straight (or even piecewise straight) in space,
then the faces must remain planar.

\begin{theorem}\label{thm:5}
  If the boundary of an uncreased flat surface $M$ is piecewise
  linear in space, then $M$ lies in a plane.
\end{theorem}
\begin{proof}
  Let $p$ be a parabolic point in the interior of $M$, a point of type (c).  We will
  show a contradiction.  It then follows that every point of $M$ is of
  type (a), (b), or (d), so planar points of type (a) are dense and by
  integration $M$ lies in a plane.

  Because $p$ is parabolic, it has by Proposition~\ref{prp:1} a
  neighborhood consisting of parabolic points which is a ruled
  surface.  By Corollary~\ref{cory:1},
  the rule segment through each point in this neighborhood can be
  extended to the boundary of $M$.  Let $U$ be the neighborhood so
  extended.

  Now the boundary of $U$ consists of a first rule segment $ab$, a last
  rule segment $cd$, and arcs $bd$ and $ac$ of which at least one must be
  nontrivial, say $bd$.  Because we extended $U$ to the boundary of
  $M$ and the boundary of $M$ is piecewise linear, $bd$ consists of a
  chain of segments.  Let $b'd'$ be one of these segments.

  Let $q$ be any point interior to the segment $b'd'$, and consider
  the normal vector $n(q)$ to $M$ at $q$.  The normal is perpendicular
  to $b'd'$ and to the rule segment $C_q$ meeting $q$.  Because $U$ is
  torsal, its derivative $n'(q)$ along $b'd'$ is also perpendicular to
  $C_q$, and because the normal is always perpendicular to $b'd'$ the
  derivative is perpendicular to $b'd'$.  But this forces $n'(q)$ to
  be a multiple of $n(q)$, therefore zero, which makes the points of
  $C_q$ planar and is a contradiction.
\end{proof}


\section{Straight Creases Stay Straight}

Next we show that straight edges of a crease pattern
must actually fold to straight line segments in space.

\begin{theorem} \label{straight creases stay straight}
  If $\gamma$ is a geodesic crease in a creased flat surface~$M$
  with fold angle distinct from $\pm 180^\circ$,
  then $\gamma$ is a segment in~$\R^3$.
\end{theorem}
\begin{proof}
  The creased surface $M$ decomposes by definition into a complex of
  uncreased surfaces, creases, and vertices.
  A point $p$ in the interior of $\gamma$ is therefore on the boundary of two
  uncreased pieces; call them $S$ and~$T$.  Let $S_p$ and $T_p$ be the tangent
  planes to $S$ and $T$ respectively at~$p$.
  Because $\gamma$ is by hypothesis not a proper semicrease,
  has no semivertices along it,
  and has a fold angle distinct from $\pm 180^\circ$,
  there is some $p \in \gamma$ where $S_p \neq T_p$.
  By continuity, the same is true for a neighborhood in $\gamma$ of~$p$;
  let $U$ be the maximal such neighborhood.

  Now parametrize $\gamma$ by arclength and let $p = \gamma(s)$.  At
  each $q = \gamma(t)$, the tangent vector $\gamma'(t)$ lies in the
  intersection $S_q \neq T_q$; in~$U$, this determines $\gamma'(t)$ up
  to sign.  Because $S$ and $T$ are $C^2$, the tangent planes $S_q$ and $T_q$
  are $C^1$, hence so is $\gamma'(t)$, and the curvature $\gamma''(t)$
  exists and is continuous.

  Now around any $q \in U$ project $\gamma$ onto the tangent plane
  $S_q$.  Because $\gamma$ is a geodesic, we get a curve of zero
  curvature at $q$, so $\gamma''(t)$ must be perpendicular to $S_q$.
  Similarly $\gamma''(t) \perp T_q$.  But certainly $\gamma''(t) \perp
  \gamma'(t)$.  So $\gamma''(t) = 0$.

  We have $\gamma''(t) = 0$ for $t$ in a neighborhood of $s$, so
  $\gamma$ is a segment on $U$.  Further, by the considerations of
  Theorem~\ref{thm:5}, the tangent planes $S_q$ and $T_q$ are constant
  on~$U$.  Therefore they remain distinct at the endpoints of~$U$, and
  because $U$ is maximal, these must be the endpoints of $\gamma$ and
  $\gamma$ is a segment.
\end{proof}

Combining the previous two theorems, we deduce that
polygonal faces of the crease pattern with no boundary edges
must indeed stay planar:

\begin{corollary}\label{planar dammit}
  If an uncreased region of a creased flat surface $M$ is piecewise
  geodesic and entirely interior to $M$, then the region lies in a plane.
\end{corollary}


\section{Nonexistence of Pleated Hyperbolic Paraboloid}

Now we can apply our theory to prove nonfoldability of crease patterns.
First we need to formally define what this means.

\begin{definition}
  A \term{piece of paper} is a planar compact 2-manifold.
  A \term{crease pattern} is a graph embedded into a piece of paper,
  with each edge embedded as a non-self-intersecting curve.
  A \term{proper folding} of a crease pattern is an isometric embedding
  of the piece of paper into 3D whose image is a good surface such that
  the union of vertices and edges of the crease pattern map onto
  the union of vertices and creases of the good surface.
  A \term{rigid folding} is a proper folding that maps each face of the
  crease pattern into a plane (and thus acts as a rigid motion on each face).
\end{definition}

Note that a proper folding must fold every edge of the crease pattern
by an angle distinct from $0$ (to be a crease) and from $\pm 180^\circ$
(to be an embedding).  We call such fold angles \term{nontrivial}.
Also, one edge of a crease pattern may map to multiple creases in 3D,
because of intervening semivertices.

The key property we need from the theory developed in the previous sections
is the following consequence of Corollary~\ref{planar dammit}:

\begin{corollary} \label{interior rigid}
  For any crease pattern made up of just straight edges,
  any proper folding must fold the interior faces rigidly.
\end{corollary}

We start by observing that the center of the standard crease pattern
for a pleated hyperbolic paraboloid has no proper folding.

\begin{lemma} \label{four triangles}
  Any crease pattern containing four right triangular faces,
  connected in a cycle along their short edges,
  has no rigid folding.
\end{lemma}

\begin{proof}
  This well-known lemma follows from,
  e.g., \cite[Lemma~9]{Donoso-O'Rourke-2002}.
  For completeness, we give a proof.
  Let $v_1$, $v_2$, $v_3$, and $v_4$ denote the direction vectors
  of the four short edges of the triangular faces, in cyclic order.
  By the planarity of the faces,
  the angle between adjacent direction vectors is kept at~$90^\circ$.
  Thus the folding angle of edge $i$ equals the angle
  between $v_{i-1}$ and $v_{i+1}$ (where indices are treated modulo~$4$).
  If edge $2$ is folded nontrivially, then
  $v_1$ and $v_3$ are nonparallel and define a single plane~$\Pi$.
  Because $v_2$ is perpendicular to both $v_1$ and~$v_3$,
  $v_2$~is perpendicular to~$\Pi$.
  Similarly, $v_4$ is perpendicular to~$\Pi$.
  Thus $v_2$ and $v_4$ are parallel,
  and hence edge $3$ is folded trivially.
  Therefore two consecutive creases cannot both be folded nontrivially.
\end{proof}

\begin{corollary} \label{hypar center}
  The standard crease pattern for a pleated hyperbolic paraboloid
  (shown in Figure~\ref{hypar cp}), with $n \geq 2$ rings,
  has no proper folding.
\end{corollary}

\begin{proof}
  With $n \geq 2$ rings, the four central triangular faces are completely
  interior.  By Corollary~\ref{interior rigid}, any proper folding
  keeps these faces planar.  But Lemma~\ref{four triangles} forbids
  these faces from folding rigidly.
\end{proof}

\begin{figure}
  \centering
  \subfigure[Holey mountain-valley pattern
             for the pleated hyperbolic paraboloid. \label{holey hypar}]
    {\includegraphics[scale=0.6]{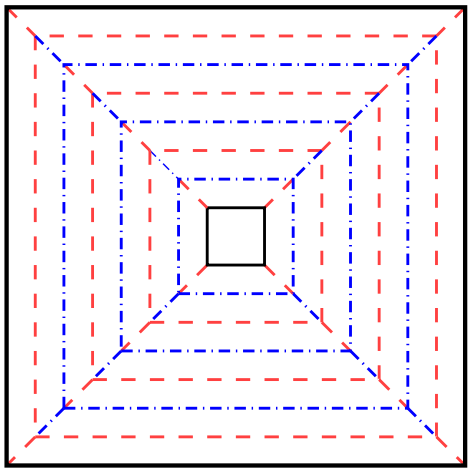}}\hfil\hfil
  \subfigure[Holey concentric pleat mountain-valley pattern.
             \label{holey concentric pleat}]
    {\includegraphics[scale=0.6]{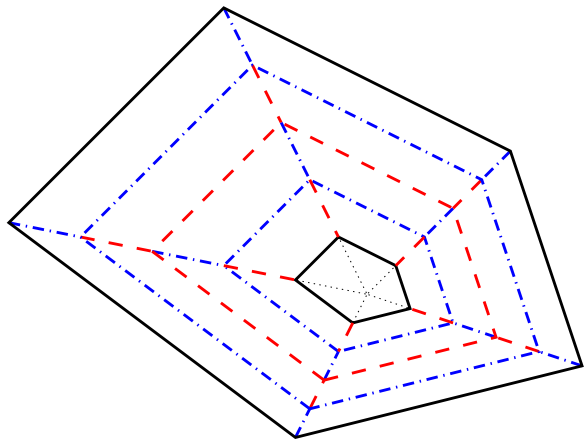}}
  \caption{Holey mountain-valley patterns which have no proper foldings.}
  \label{holey}
\end{figure}

The standard crease pattern for a pleated hyperbolic paraboloid
cannot fold properly for a deeper reason than the central ring.
To prove this, we consider the \term{holey crease pattern}
in which the central ring of triangles has been cut out,
as shown in Figure~\ref{holey hypar}.
If there were $n$ rings in the initial crease pattern
(counting the central four triangles as one ring), then $n-1$ rings remain.

\begin{theorem}
  The holey crease pattern for a pleated hyperbolic paraboloid
  (shown in Figure~\ref{holey hypar}), with $n-1 \geq 3$ rings,
  has no proper folding.
\end{theorem}

\begin{proof}
  Consider any nonboundary square ring of the crease pattern.
  By Corollary~\ref{interior rigid}, the four trapezoidal faces
  each remain planar.  Any folding of these four faces in fact
  induces a folding of their extension to four meeting right triangles.
  But Lemma~\ref{four triangles} forbids these faces from folding rigidly.
\end{proof}

A different argument proves nonfoldability of a more general
pleated crease pattern.  
Define the \term{concentric pleat crease pattern}
to consist of $n$ uniformly scaled copies of a convex polygon~$P$,
in perspective from an interior point~$p$, together with the ``diagonals''
connecting $p$ to each vertex of each copy of $P$.
The outermost copy of the polygon $P$ is the boundary of the piece of paper,
and in the \term{holey concentric pleat mountain-valley pattern}
we additionally cut out a hole bounded by the innermost copy of~$P$.
Thus $n-1$ rings remain;
Figure~\ref{holey concentric pleat} shows an example.

First we need to argue about which creases can be mountains and valleys.

\begin{definition}
  A \term{mountain-valley pattern} is a crease pattern together with
  an assignment of signs ($+1$ for ``mountain'' and $-1$ for ``valley'')
  to the edges of a crease pattern.
  A \term{proper folding} of a mountain-valley pattern,
  in addition to being a proper folding of the crease pattern,
  must have the signs of the fold angles (relative to some canonical
  orientation of the top side of the piece of paper) match the signs of the
  mountain-valley assignment.
\end{definition}

\begin{lemma} \label{not all mountains}
  A single-vertex mountain-valley pattern consisting of entirely mountains
  or entirely valleys has no proper rigid folding.
\end{lemma}

\begin{proof}
  If we intersect the piece of paper with a (small) unit sphere centered
  at the vertex, we obtain a spherical polygon whose edge lengths match the
  angles between consecutive edges of the crease pattern.
  (Here we rely on the fact that the vertex is intrinsically flat,
   so that the polygon lies in a hemisphere and thus no edges go
   the ``wrong way'' around the sphere.)
  The total perimeter of the spherical polygon is $360^\circ$.
  Any rigid folding induces a non-self-intersecting spherical polygon,
  with mountain folds mapping to convex angles and
  valley folds mapping to reflex angles, or vice versa
  (depending on the orientation of the piece of paper).
  To be entirely mountain or entirely valley,
  the folded spherical polygon must be locally convex,
  and by non-self-intersection, convex.
  But any convex spherical polygon (that is not planar) has perimeter
  strictly less than $360^\circ$ \cite[page~265, Theorem~IV]{Halsted-1885},
  a contradiction.
\end{proof}

\begin{lemma} \label{degree 4}
  Consider a rigidly foldable degree-$4$ single-vertex mountain-valley
  pattern with angles $\theta_1$, $\theta_2$, $\theta_3$, and $\theta_4$
  in cyclic order.
  Then exactly one edge of the mountain-valley pattern has sign different
  from the other three, and if $\theta_2 + \theta_3 \geq 180^\circ$,
  then the unique edge cannot be the one between $\theta_2$ and~$\theta_3$.
\end{lemma}

\begin{proof}
  Again we intersect the piece of paper with a (small) unit sphere centered
  at the vertex to obtain a spherical polygon, with edge lengths $\theta_1$,
  $\theta_2$, $\theta_3$, and $\theta_4$, and whose convex angles correspond
  to mountain folds and whose reflex angles correspond to valley folds,
  or vice versa.  By Lemma~\ref{not all mountains}, at least one vertex
  is reflex, and thus the remaining vertices must be convex.
  (The only non-self-intersecting spherical polygons with only two convex
   vertices lie along a line, and hence have no reflex vertices.
   Here we rely on the fact that the vertex is intrinsically flat,
   so that the polygon lies in a hemisphere, to define the interior.)
  The two edges incident to this vertex form a triangle,
  by adding a closing edge.
  The other two edges of the quadrilateral also form a triangle,
  with the same closing edge, that strictly contains the previous triangle.
  The latter triangle therefore has strictly larger perimeter than the
  former triangle, as any convex spherical polygon has larger perimeter than
  that of any convex spherical polygon it contains
  \cite[page~264, Theorem~III]{Halsted-1885}.
  The two triangles share an edge which appears in both perimeter sums,
  so we obtain that the two edges incident to the reflex angle sum to less
  than half the total perimeter of the quadrilateral, which is $360^\circ$.
  Therefore they cannot be the edges corresponding to angles
  $\theta_2$ and~$\theta_3$.
\end{proof}

Now we can prove the nonfoldability of a general concentric pleat:

\begin{theorem}
  The holey concentric pleat crease pattern
  (shown in Figure~\ref{holey concentric pleat}), with $n-1 \geq 4$ rings,
  has no proper folding.
\end{theorem}

\begin{proof}
  First we focus on two consecutive nonboundary rings of the crease pattern,
  which by Corollary~\ref{interior rigid} fold rigidly.
  Each degree-$4$ vertex between the two rings
  has a consecutive pair of angles summing to more than $180^\circ$
  (the local exterior of~$P$), and two consecutive pairs of angles summing to
  exactly $180^\circ$ (because the diagonals are collinear).
  By Lemma~\ref{degree 4}, the interior diagonal must be the unique crease
  with sign different from the other three.
  Thus all of the creases between the rings have the same sign,
  which is the same sign as all of the diagonal creases in the outer ring.

  Now focus on the outer ring, whose diagonal creases all have the same sign.
  Any folding of the faces of a ring in fact induces a folding of their
  extension to meeting triangles.
  (In the unfolded state, the central point is the center $p$ of scaling.)
  Thus we obtain a rigid folding of a crease pattern with a single vertex $p$
  and one emanating edge per vertex of~$P$, all with the same sign.
  But such a folding contradicts Lemma~\ref{not all mountains} .
\end{proof}


\section{Existence of Triangulated Hyperbolic Paraboloid}

\begin{figure}
  \centering
  \subfigure[Asymmetric triangulation. \label{asymmetric triangulation}]
            {\includegraphics[scale=0.6]{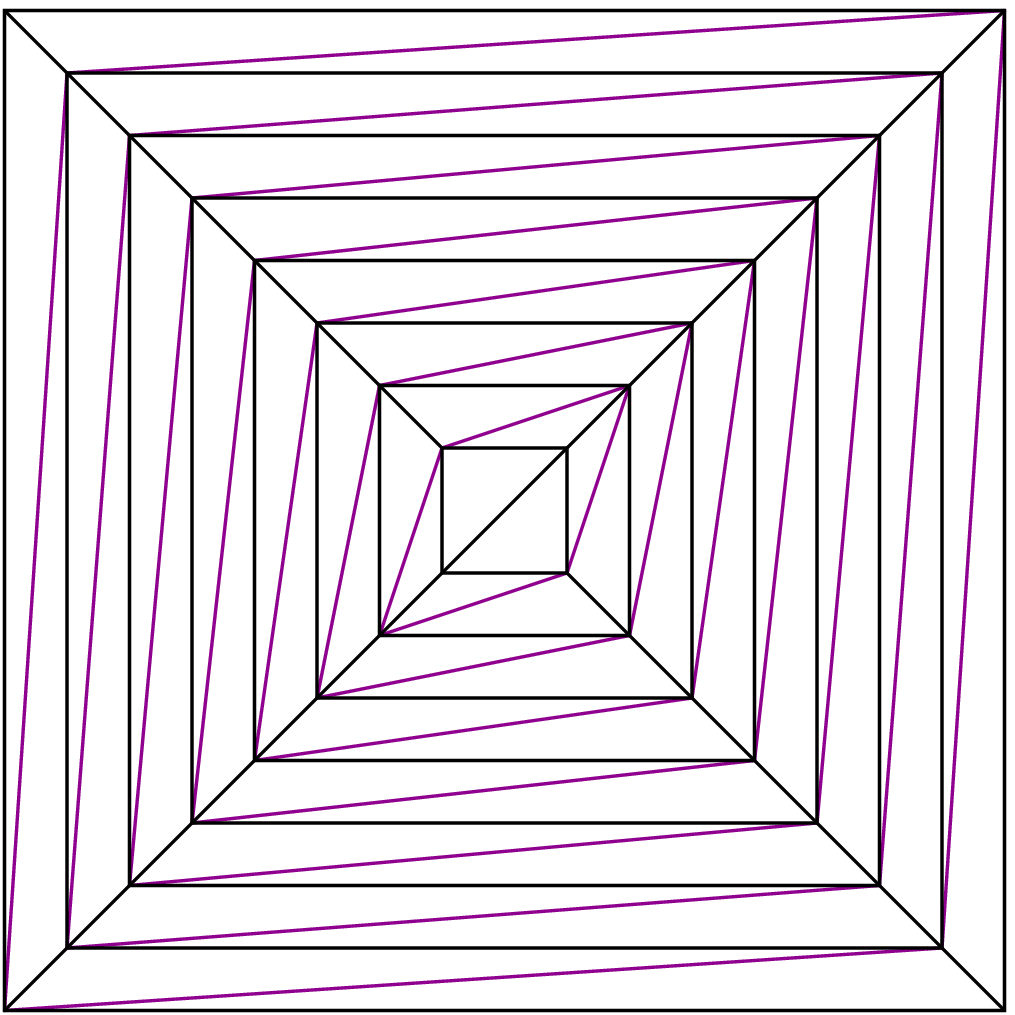}}
  \hfil
  \subfigure[Alternating asymmetric triangulation.
             \label{alternating asymmetric triangulation}]
            {\includegraphics[scale=0.6]{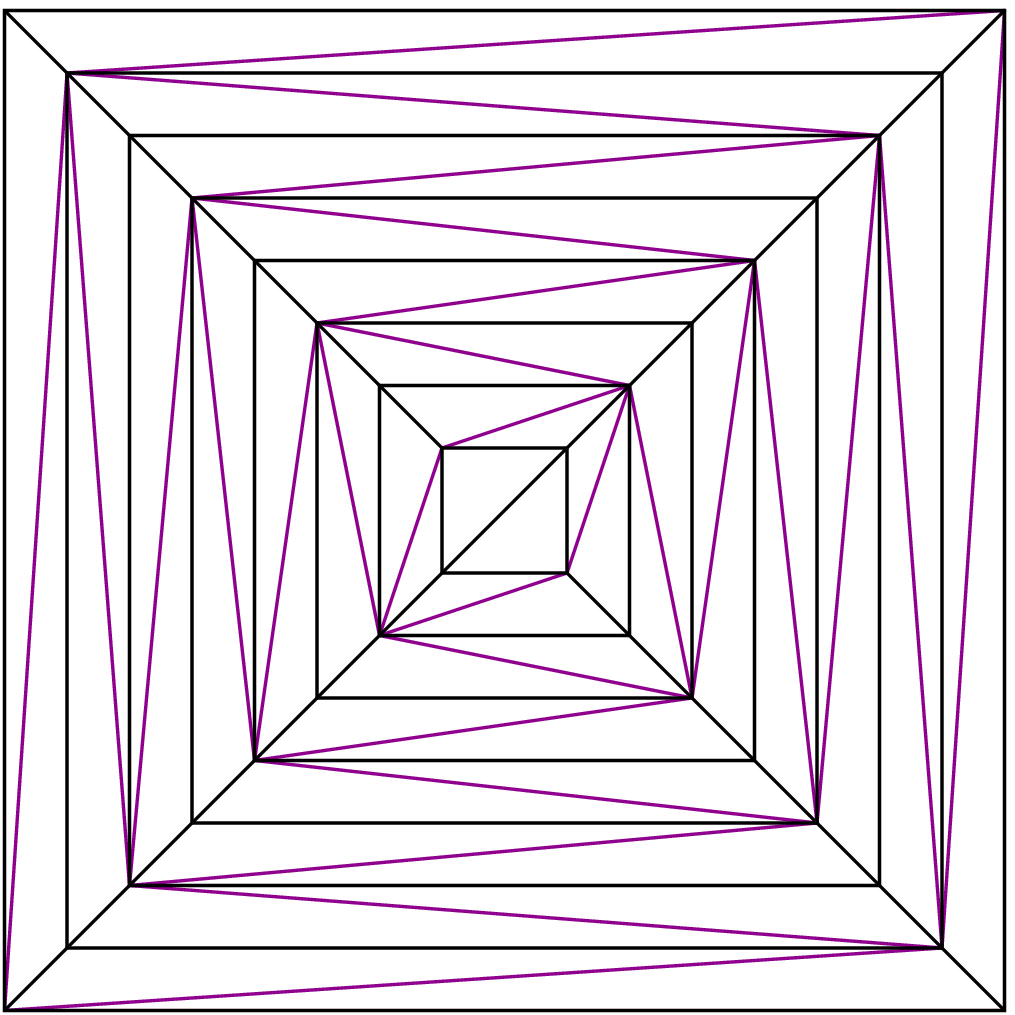}}
  \caption{Two foldable triangulations of the hyperbolic paraboloid
           crease pattern (less one diagonal in the center).}
  \label{hypar triangulations}
\end{figure}

In contrast to the classic hyperbolic paraboloid model,
we show that triangulating each trapezoidal face
and retriangulating the central ring permits folding:

\begin{theorem}
  The two mountain-valley patterns in Figure~\ref{hypar triangulations},
  with mountains and valleys matching the hyperbolic paraboloid
  of Figure~\ref{hypar mv}, have proper foldings,
  uniquely determined by the fold angle $\theta$ of the central diagonal,
  that exist at least for $n = 100$ and
  $\theta \in \{2^\circ,4^\circ,6^\circ,\dots,178^\circ\}$
  for the alternating asymmetric triangulation,
  and at least for $n$ and $\theta$ shown in Table~\ref{n vs theta}
  for the asymmetric triangulation.
  For each $\theta \in \{2^\circ,4^\circ,\dots,178^\circ\}$,
  the asymmetric triangulation has no proper folding for $n$ larger
  than the limit values shown in Table~\ref{n vs theta}.
\end{theorem}

\begin{table}
  \centering
  \begin{tabular}{c|cccccccccccccccccccc}
    $\theta$ & $2^\circ$ & $4^\circ$ & $6^\circ$ & $8^\circ$ & $10^\circ$ & $12^\circ$ & $14^\circ$ & $16^\circ$ & $18^\circ$ & $20^\circ$ & $22^\circ$ & $24^\circ$ & $26^\circ$ & $28^\circ$ & $30^\circ$ & $32^\circ$
    \\ \hline
    $n$ & $133$ & $67$ & $45$ & $33$ & $27$ & $23$ & $19$ & $17$ & $15$ & $13$ & $13$ & $11$ & $11$ & $9$ & $9$ & $9$
    \bigskip \\
    $\theta$ & $34^\circ$ & $36^\circ$ & $38^\circ$ & $40^\circ$ & $42^\circ$ & $44^\circ$ & $46^\circ$ & $48^\circ$ & $50^\circ$ & $\cdots$ & $72^\circ$ & $74^\circ$ & $76^\circ$ & $\cdots$ & $176^\circ$ & $178^\circ$
    \\ \hline
    $n$ & $9$ & $7$ & $7$ & $7$ & $7$ & $7$ & $7$ & $5$ & $5$ & $\cdots$ & $5$ & $3$ & $3$ & $\cdots$ & $3$ & $3$
  \end{tabular}
  \caption{The largest $n$ for which the asymmetric triangulation has a proper
    folding, for each $\theta \in \{2^\circ, 4^\circ, \dots, 178^\circ\}$.
    (By contrast, the alternating asymmetric triangulation
     has a proper folding for $n=100$ for all such~$\theta$.)
    Interestingly, for $\theta$ not too large,
    $n \cdot \theta$ is roughly $270^\circ$.}
  \label{n vs theta}
\end{table}

\begin{figure}
  \centering
  \subfigure[Asymmetric triangulation, $\theta = 8^\circ$, $n=16$.]
    {\quad\includegraphics[scale=0.3,clip,trim=15 260 10 255]{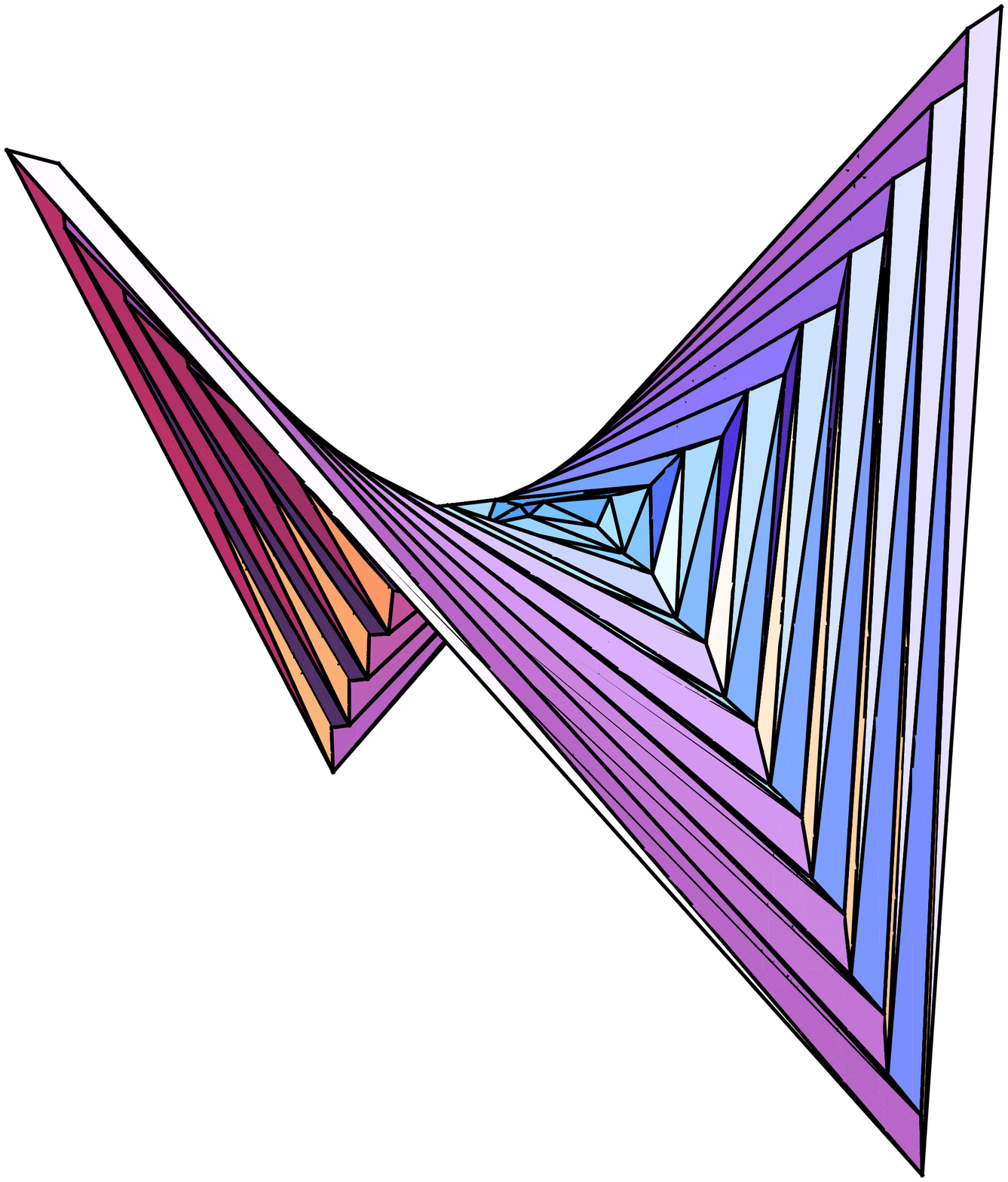}\quad}\hfil
  \subfigure[Alternating asymmetric triangulation, $\theta = 30^\circ$, $n=16$.
             \label{example alternating}]
    {\includegraphics[scale=0.4,clip,trim=130 20 130 20]{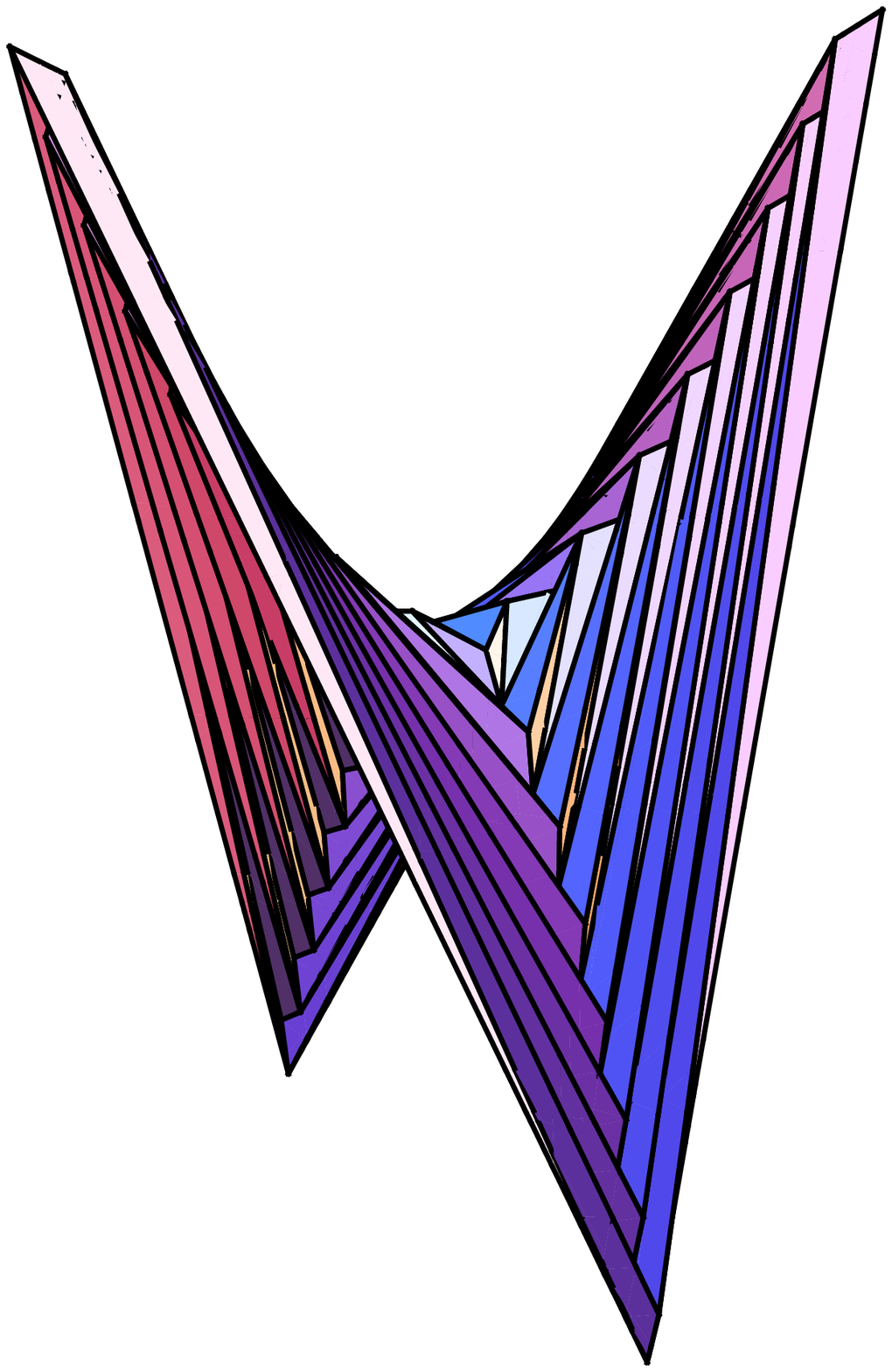}}
  \caption{Proper foldings of triangulated hyperbolic paraboloids.}
  \label{triangulated hypar}
\end{figure}

\begin{proof}
  The proof is by construction: we give a construction which implies
  uniqueness, and then use the resulting algorithm to construct the explicit
  3D geometry using interval arithmetic and a computer program.

  To get started, we are given the fold angle $\theta$ between the two
  triangles of the central square.  By fixing one of these triangles in
  a canonical position, we obtain the coordinates of the central square's
  vertices by a simple rotation.

  We claim that all other vertices are then determined by a sequence of
  intersection-of-three-spheres computations from the inside out.
  In the asymmetric triangulation of Figure~\ref{asymmetric triangulation},
  the lower-left and upper-right corners of each square have three known
  (creased) distances to three vertices from the previous square.
  Here we use Theorem~\ref{straight creases stay straight} which guarantees that
  the creases remain straight and thus their endpoints have known distance.
  Thus we can compute these vertices as the intersections of three spheres
  with known centers and radii.  Afterward, the lower-right and upper-left
  corners of the same square have three known (creased) distances to three
  known vertices, one from the previous square and two from the current square.
  Thus we can compute these vertices also as the intersection of three spheres.
  In the alternating asymmetric triangulation of
  Figure~\ref{asymmetric triangulation}, half of the squares behave the same,
  and the other half compute their corners in the opposite order
  (first lower-right and upper-left, then lower-left and upper-right).

  The intersection of three generic spheres is zero-dimensional, but in general
  may not exist and may not be unique.  Specifically, the intersection of
  two distinct spheres is either a circle, a point, or nothing;
  the further intersection with another sphere whose center is not
  collinear with the first two spheres' centers is either two points,
  one point, or nothing.  When there are two solutions, however,
  they are reflections of each other through the plane containing the
  three sphere centers.  (The circle of intersection of the first two
  spheres is centered at a point in this plane, and thus the two
  intersection points are equidistant from the plane.)

  For the hyperbolic paraboloid, we can use the mountain-valley assignment
  (from Figure~\ref{hypar mv}) to uniquely determine which intersection to
  choose.  In the first intersection of three spheres, one solution would make
  two square creases mountains (when the solution is below the plane) and the
  other solution would make those creases valleys.  Thus we choose whichever
  is appropriate for the alternation.  In the second intersection of three
  spheres, one solution would make a diagonal crease mountain, and the other
  solution would make that crease valley.  Again we choose whichever is
  appropriate for the alternation.  Therefore the folding is uniquely
  determined by $\theta$ and by the mountain-valley assignment of the
  original hyperbolic paraboloid creases.

  This construction immediately suggests an algorithm to construct the
  proper folding.  The coordinates of intersection of three spheres can be
  written as a radical expression in the center coordinates and radii
  (using addition, subtraction, multiplication, division, and square roots).
  See \cite{Trilateration-wiki} for one derivation;
  Mathematica's fully expanded solution for the general case
  (computed with \texttt{Solve}) uses over 150,000 leaf expressions
  (constant factors and variable occurrences).
  Thus, if the coordinates of the central square can be represented
  by radical expressions (e.g., $\theta$ is a multiple of $15^\circ$),
  then all coordinates in the proper folding can be so represented.
  Unfortunately, we found this purely algebraic approach to be
  computationally infeasible beyond the second square; the expressions
  immediately become too unwieldy to manipulate (barring some
  unknown simplification which Mathematica could not find).

  Therefore we opt to approximate the solution, while still guaranteeing
  that an exact solution exists, via \emph{interval arithmetic}
  \cite{Hayes-2003-interval,Moore-Kearfott-Cloud-2009,Alefeld-Herzberger-1983}.
  The idea is to represent every coordinate $x$ as an interval $[x_L,x_R]$
  of possible values, and use conservative estimates in every arithmetic
  operation and square-root computation to guarantee that the answer
  is in the computed interval.
  For example, $[a,b]+[c,d] = [a+c,b+d]$ and
  $[a,b] \cdot [c,d] = [\min\{a \cdot c, a \cdot d, b \cdot c, b \cdot d\},
  \max\{a \cdot c, a \cdot d, b \cdot c, b \cdot d\}]$,
  while $\sqrt{[a,b]}$ requires a careful implementation of a square-root
  approximation algorithm such as Newton's Method.
  The key exception is that $\sqrt{[a,b]}$ is undefined when $a < 0$.
  A negative square root is the only way that the intersection of three
  spheres, and thus the folding, can fail to exist.
  If we succeed in computing an approximate folding using interval
  arithmetic without attempting to take the square root of a partially
  negative interval, then an exact folding must exist.
  Once constructed, we need only check that the folding does not
  intersect itself (i.e., forms an embedding).

  We have implemented this interval-arithmetic construction in Mathematica;
  refer to Appendix~\ref{mathematica}.
  Using sufficiently many (between 1,024 and 2,048)
  digits of precision in the interval computations,
  the computation succeeds for the claimed ranges of $n$ and~$\theta$
  for both triangulations.
  Table~\ref{precision} shows how the required precision grows with~$n$
  (roughly linearly), for a few different instances.
  Figure~\ref{triangulated hypar} shows some of the computed structures
  (whose intervals are much smaller than the drawn line thickness).
  The folding construction produces an answer
  for the asymmetric triangulation even for $n=100$ and
  $\theta \in \{2^\circ, 4^\circ, \dots, 178^\circ\}$,
  but the folding self-intersects for $n$ larger
  than the limit values shown in Table~\ref{n vs theta}.
\end{proof}

\begin{table}
  \centering
  \begin{tabular}{l|rrrrrrrr}
    digits of precision & $16$ & $32$ & $64$ & $128$ & $256$ & $512$ & $1024$ & $2048$
    \\ \hline
    $n$ for $\theta=1^\circ$ &  $3$ &  $6$ & $12$ &  $22$ &  $41$ &  $76$ & $\geq 100$
    \\
    $n$ for $\theta=1^\circ$ alt.&  $3$ &  $6$ & $12$ &  $24$ &  $43$ &  $79$ & $\geq 100$
    \\
    $n$ for $\theta=45^\circ$ alt.&  $3$ &  $5$ & $10$ &  $18$ &  $32$ &  $58$ & $\geq 100$
    \\
    $n$ for $\theta=76^\circ$ alt.&  $2$ &  $5$ &  $9$ &  $16$ &  $29$ &  $53$ & $95$ & $\geq 100$
  \end{tabular}
  \caption{Number $n$ of triangulated rings that can be successfully
    constructed using various precisions (measured in digits) of interval
    arithmetic.}
  \label{precision}
\end{table}

We conjecture that this theorem holds for all $n$ and all
$\theta < 180^\circ$ for the alternating asymmetric triangulation,
but lack an appropriately general proof technique.
If the construction indeed works for all $\theta$ in some interval
$[0,\Theta)$, then we would also obtain a continuous folding motion.

\begin{figure}
  \centering
  \begin{minipage}{1.9in}
    \subfigure[Actual zig-zag and parabolic fits.]
              {\quad\includegraphics[width=1.8in]{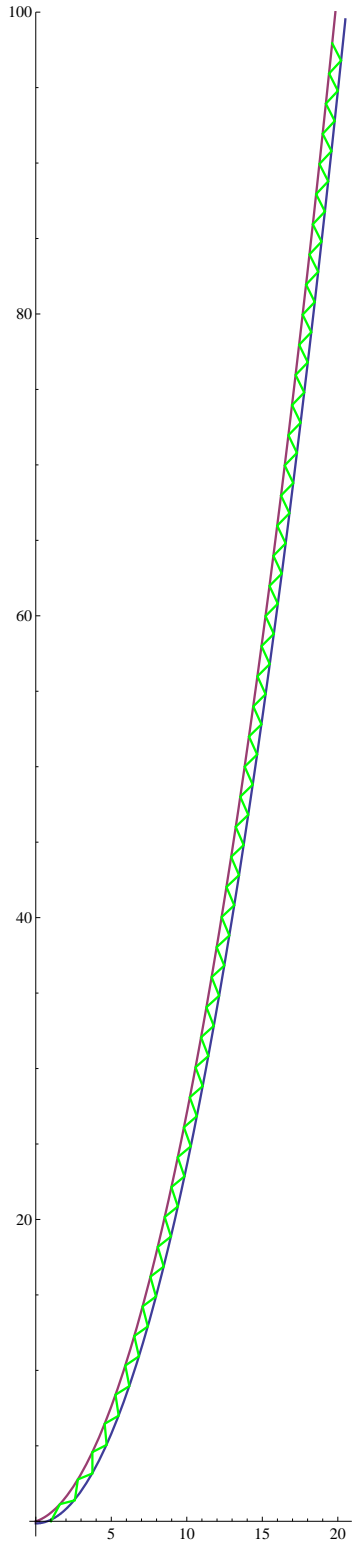}\quad}
  \end{minipage}\hfill
  \begin{minipage}{4in}
    \subfigure[Absolute difference: fit minus actual.]
              {\includegraphics[width=4in]{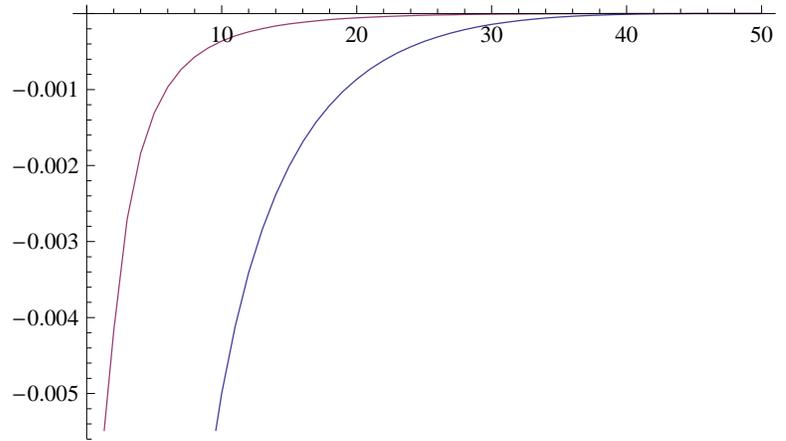}}
    \subfigure[Relative difference: fit over actual.]
              {\includegraphics[width=4in]{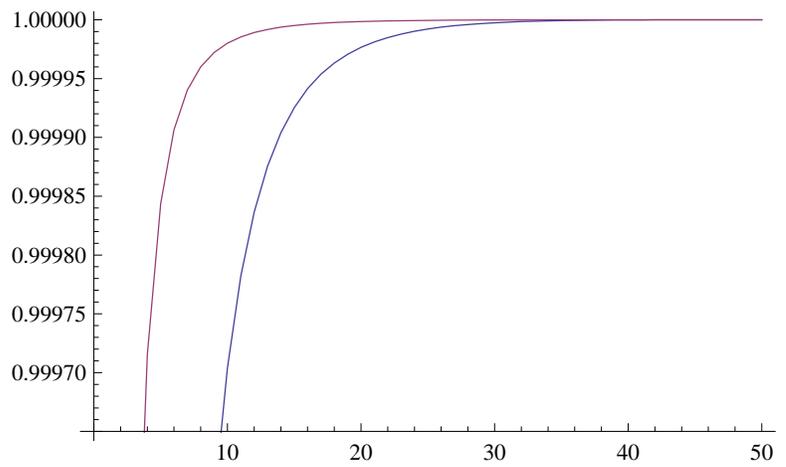}}
  \end{minipage}
  \caption{Planar cross-section of alternating asymmetric triangulation,
           $\theta = 30^\circ$, $n=100$, with parabolic fits of each parity
           class based on the last three vertices.}
  \label{zigzagall}
\end{figure}

Interestingly, the diagonal cross-sections of these structures seem to
approach parabolic in the limit.  Figure~\ref{zigzagall} shows the
$x = y \geq 0$ cross-section of the example from
Figure~\ref{example alternating}, extended out to $n=100$.
The parabolic fit we use for each parity class is the unique quadratic
polynomial passing through the three points in the parity class farthest
from the center.  The resulting error near the center is significant, but
still much smaller than the diagonal crease length,~$\sqrt 2$.
Least-square fits reduce this error
but do not illustrate the limiting behavior.

\section{Smooth Hyperbolic Paraboloid}

Given a smooth plane curve $\Gamma$ and an embedding of $\Gamma$ in
space as a smooth space curve $\gamma$, previous work
\cite{FuchsTabachnikov} has studied the problem of folding a strip of
paper so that a crease in the form of $\Gamma$ in the plane follows
the space curve $\gamma$ when folded.  The main theorem from this work
is that such a folding always exists, at least for a sufficiently
narrow strip about $\Gamma$, under the condition that the curvature of
$\gamma$ be everywhere strictly greater than that of $\Gamma$.

Further, with some differential geometry described in
\cite{FuchsTabachnikov}, it is possible to write down exactly how the
strip folds in space; there are always exactly two possible choices,
and additionally two ways to fold the strip so that $\Gamma$ lies
along $\gamma$ but remains uncreased.

Based on some preliminary work using these techniques, we conjecture
that the circular pleat indeed folds, and that so too does any similar crease
pattern consisting of a concentric series of convex smooth curves.
Unfortunately a proof remains elusive.  Such a proof would be the first
proof to our knowledge of the existence of any curved-crease origami model,
beyond the local neighborhood of a single crease.


\section*{Acknowledgments}

We thank sarah-marie belcastro, Thomas Hull, and Ronald Resch for
helpful related discussions over the years.
We also thank Jenna Fizel for folding and photographing the models
in Figures~\ref{hypar photo} and~\ref{circular photo}.


\bibliography{paper}
\bibliographystyle{alpha}


\appendix
\section{Mathematica Construction of Triangulated Hyperbolic Paraboloid}
\label{mathematica}

\begin{small}
\begin{verbatim}
SW = 1; SE = 2; NE = 3; NW = 4;
next[d_] := 1 + Mod[d, 4];
prev[d_] := 1 + Mod[d + 2, 4];
square[i_] := {{-i, -i}, {i, -i}, {i, i}, {-i, i}};
side[i_] := 2 i;
diagonal[i_] := Sqrt[2] side[i];
diagpiece[i_] := Sqrt[2];
cross[i_] := Sqrt[(i - (i - 1))^2 + (i + (i - 1))^2];
n = 100;
bigside := side[n];
bigdiagonal := diagonal[n];
trapezoid[i_, d_] := 
  square[i][[{d, next[d]}]] ~Join~ square[i - 1][[{next[d], d}]];

(* Note: \[Theta\] is half the fold angle. *)
algebraicmiddle := {
  {-1, -1, 0}, {-Cos[\[Theta]], Cos[\[Theta]], Sqrt[2] Sin[\[Theta]]},
  {1, 1, 0}, {Cos[\[Theta]], -Cos[\[Theta]], Sqrt[2] Sin[\[Theta]]}};
precision = 1024;
middle := Map[Interval[N[#, precision]] &, algebraicmiddle, {2}];

orientation[a_, b_, c_, d_] := 
  Sign[Det[Append[#, 1] & /@ {a, b, c, d}]];
pickorientation[choices_, a_, b_, c_, pos_] :=
  Module[
    {select = Select[choices, 
                     If[pos, orientation[a, b, c, #] >= 0 &, 
                             orientation[a, b, c, #] <= 0 &]]},
    If[Length[select] > 1, 
      Print["Warning: Multiple choices in pickorientation..."]];
    If[Length[select] > 0, select[[1]], 
      Print["Couldn't find mountain/valley choice"]]];
triangles[1] := {triangulated[[1, {1, 2, 3}]], triangulated[[1, {3, 4, 1}]]};

threespheres = 
  Solve[{SquaredEuclideanDistance[{x1, y1, z1}, {x, y, z}] == d1^2,
    SquaredEuclideanDistance[{x2, y2, z2}, {x, y, z}] == d2^2,
    SquaredEuclideanDistance[{x3, y3, z3}, {x, y, z}] == d3^2}, {x, y, z}];
solvethreespheres[{X1_, Y1_, Z1_}, D1_, {X2_, Y2_, Z2_}, 
   D2_, {X3_, Y3_, Z3_}, D3_] := 
  threespheres /. {x1 -> X1, y1 -> Y1, z1 -> Z1, x2 -> X2, y2 -> Y2, 
    z2 -> Z2, x3 -> X3, y3 -> Y3, z3 -> Z3, d1 -> D1, d2 -> D2, d3 -> D3};

checklengths := Module[{i, j, t, tri, error, maxerror, who},
  maxerror = 0;
  who = "I did not see any error!";
  For[i = 1, i <= Length[triangulated], i++,
    For[t = 1, t <= Length[triangles[i]], t++,
      tri = triangles[i][[t]];
      For[j = 1, j <= 3, j++,
        error = 
          Min[Abs[EuclideanDistance[tri[[j]], tri[[Mod[j, 3] + 1]]] - #] & /@ 
              If[i == 1, {side[i], 2 diagpiece[i]},
                         {side[i], side[i - 1], cross[i], diagpiece[i]}]];
        If[error >= maxerror,
          who = {"i=", i, " triangle ", t, " edge ", j, "-", 
            Mod[j, 3] + 1, " has error ", error, " = ", N[error], ": ", 
            EuclideanDistance[tri[[j]], tri[[Mod[j, 3] + 1]]], 
            " vs. ", {side[i], side[i - 1], cross[i], diagpiece[i]}};
          maxerror = error]]]];
  Print @@ who];

(* Collision detection *)
stL = Solve[{(1 - L) ax + L bx == 
        px + s (qx - px) + t (rx - px), (1 - L) ay + L by == 
        py + s (qy - py) + t (ry - py), (1 - L) az + L bz == 
        pz + s (qz - pz) + t (rz - pz)}, {s, t, L}][[1]];
pierces[seg_, tri_] := Module[{parms},
  Check[Quiet[
    parms = stL /. {ax -> seg[[1]][[1]], ay -> seg[[1]][[2]], 
       az -> seg[[1]][[3]], bx -> seg[[2]][[1]], by -> seg[[2]][[2]],
        bz -> seg[[2]][[3]], px -> tri[[1]][[1]], 
       py -> tri[[1]][[2]], pz -> tri[[1]][[3]], qx -> tri[[2]][[1]],
        qy -> tri[[2]][[2]], qz -> tri[[2]][[3]], 
       rx -> tri[[3]][[1]], ry -> tri[[3]][[2]], 
       rz -> tri[[3]][[3]]};
    spt = (s + t) /. parms; LL = L /. parms;
    Return[(s > 0 && t > 0 && s + t < 1 && 0 < L < 1) /. 
      parms], {Power::"infy", \[Infinity]::"indet"}], 
   Return[False], {Power::"infy", \[Infinity]::"indet"}]];
checkcollision := Module[{i, j, t, t2, tri, seg},
  For[i = 1, i <= Length[triangulated] - 1, i++,
    PrintTemporary["Ring ", i, " vs. ", i + 1];
    For[t = 1, t <= Length[triangles[i]], t++,
      tri = triangles[i][[t]];
      For[t2 = 1, t2 <= Length[triangles[i + 1]], t2++,
        For[j = 1, j <= 3, j++,
         seg = triangles[i + 1][[t2]][[{j, Mod[j, 3] + 1}]];
         If[Length[Intersection[seg, tri]] > 0, Continue[]];
         If[pierces[seg, tri],
           Print["COLLISION! ", i, ",", t, " vs. ", i + 1, ",", t2, ",", j];
           Return[{seg, tri}]]]]]];
  Print["No collision."]];

adaptiveprecisiontest[mint_, maxt_, tstep_: 1, minprecision_: 8, maxprecision_: 65536] := 
  Module[{t},
    precision = minprecision;
    Print["n: ", n];
    For[t = mint, t <= maxt, t = t + tstep,
      Print[t, " degrees:"];
      \[Theta] = t \[Pi]/180;
      For[True, precision <= maxprecision, precision = 2 precision,
        Print["precision: ", precision];
        If[computetriangulated == n,
          Print["Required precision for ", t, " degrees:", precision];
          checklengths; checkcollision; Break[]]]]];

Print["ALTERNATING ASYMMETRIC TRIANGULATED FOLDING"];

oddoutset [i_, d_, last_] := 
  pickorientation[{x, y, z} /. # & /@ 
    solvethreespheres[last[[d]], diagpiece[i], last[[next[d]]], 
      cross[i], last[[prev[d]]], cross[i]],
    last[[prev[d]]], last[[d]], last[[next[d]]], OddQ[i]];
evenoutset[i_, d_, last_, odds_] := 
  pickorientation[{x, y, z} /. # & /@ 
    solvethreespheres[last[[d]], diagpiece[i], odds[[next[d]]], 
      side[i], odds[[prev[d]]], side[i]],
    odds[[prev[d]]], last[[d]], odds[[next[d]]], OddQ[i]];
computetriangulated := Module[{i, odds},
  triangulated = {middle};
  For[i = 2, i <= n, i++,
    If[OddQ[i],
      last = RotateRight[triangulated[[-1]]],
      last = triangulated[[-1]]];
    odds = Table[If[OddQ[d], oddoutset[i, d, last]], {d, 1, 4}];
    Print["Round ", i, " odds done"];
    If[Count[odds, Null] > 2, Break[]];
    both = 
      Table[If[OddQ[d], odds[[d]], evenoutset[i, d, last, odds]], {d, 1, 4}];
    If[OddQ[i],
      AppendTo[triangulated, RotateLeft[both]],
      AppendTo[triangulated, both]];
    Print["Round ", i, " evens done"];
    If[Count[triangulated[[-1]], Null] > 0, Break[]]];
  Print[i - 1, " rounds complete"];
  i-1];
triangles[i_] := 
  Flatten[Table[
    If[OddQ[d] == EvenQ[i],
      {{triangulated[[i, d]], triangulated[[i - 1, d]], 
        triangulated[[i - 1, prev[d]]]},
       {triangulated[[i, d]], triangulated[[i - 1, d]], 
        triangulated[[i - 1, next[d]]]}},
      {{triangulated[[i, d]], triangulated[[i - 1, d]], 
        triangulated[[i, prev[d]]]},
       {triangulated[[i, d]], triangulated[[i - 1, d]],
        triangulated[[i, next[d]]]}}], {d, 1, 4}], 1];

adaptiveprecisiontest[1, 90];

Print["ASYMMETRIC TRIANGULATED FOLDING"];

computetriangulated := Module[{i, odds},
  triangulated = {middle};
  For[i = 2, i <= n, i++,
    odds = Table[If[OddQ[d], oddoutset[i, d, triangulated[[-1]]]], {d, 1, 4}];
    Print["Round ", i, " odds done!"];
    If[Count[odds, Null] > 2, Break[]];
    AppendTo[triangulated, 
      Table[If[OddQ[d], odds[[d]], 
        evenoutset[i, d, triangulated[[-1]], odds]], {d, 1, 4}]];
    Print["Round ", i, " evens done"];
    If[Count[triangulated[[-1]], Null] > 0, Break[]]];
  Print[i - 1, " rounds complete"];
  i-1];
triangles[i_] := 
  Flatten[Table[
    If[OddQ[d],
      {{triangulated[[i, d]], triangulated[[i - 1, d]], 
        triangulated[[i - 1, prev[d]]]},
       {triangulated[[i, d]], triangulated[[i - 1, d]], 
        triangulated[[i - 1, next[d]]]}},
      {{triangulated[[i, d]], triangulated[[i - 1, d]], 
        triangulated[[i, prev[d]]]},
       {triangulated[[i, d]], triangulated[[i - 1, d]],
        triangulated[[i, next[d]]]}}], {d, 1, 4}], 1]

adaptiveprecisiontest[1, 90];
\end{verbatim}
\end{small}

\end{document}